\documentclass[%
 reprint,
superscriptaddress,
 amsmath,amssymb,
 aps, prl
]{revtex4-1}
\usepackage{graphicx}
\usepackage{amsfonts}
\usepackage{dcolumn}
\usepackage{bm}
\usepackage{hyperref}

\newcommand{\beq}{\begin{equation}}
\newcommand{\eeq}{\end{equation}}
\newcommand{\beqa}{\begin{eqnarray}}
\newcommand{\eeqa}{\end{eqnarray}}

\makeatletter
\newcommand{\doublewidetilde}[1]{{%
  \mathpalette\double@widetilde{#1}%
}}
\newcommand{\double@widetilde}[2]{%
  \sbox\z@{$\m@th#1\widetilde{#2}$}%
  \ht\z@=.9\ht\z@
  \widetilde{\box\z@}%
}
\makeatother

\usepackage{calrsfs}
\DeclareMathAlphabet{\pazocal}{OMS}{zplm}{m}{n}

\begin{document}

\title{Vanishing efficiency of speeded-up quantum Otto engines}
\author{A. Tobalina}
\affiliation{Department of Physical Chemistry, University of the Basque Country UPV/EHU,\\ Apdo 644, Bilbao, Spain}
\author{I. Lizuain}
\affiliation{Department of Applied Mathematics, University of the Basque Country UPV/EHU,
Plaza Europa 1, 20018 Donostia-San Sebastian, Spain}
\author{J. G. Muga}
\affiliation{Department of Physical Chemistry, University of the Basque Country UPV/EHU,\\ Apdo 644, Bilbao, Spain}


\begin{abstract}
We assess the energy cost of shortcuts to adiabatic expansions or compressions of a  
harmonic oscillator, the power strokes of a quantum Otto engine. 
Difficulties to identify the cost stem from the interplay between different parts of the total system (the primary system -the particle-,  and the control system) 
and definitions of work (exclusive and inclusive). While attention is usually paid to the inclusive work of the primary system,
we identify the energy cost as the exclusive work of the 
total system, which, for a  clearcut scale disparity
between primary and control systems,  
coincides with the exclusive work for the control system. This energy cost 
makes the efficiency of the engine zero. 
For an ion in a Paul trap, the Paul trap fixes the  gauge for the primary system, resulting in a counterintuitive evolution of its inclusive power and internal energy. Conditions for which inclusive power of the primary system and exclusive power control system are proportional are found. 
\end{abstract}
\maketitle
%
%
%
%
%
%
%
%
%
%
%
{\it Introduction.} One of the  challenges to realize 
\textit{quantum technologies} and devices that outperform classical counterparts,
is to achieve exhaustive control over the state and dynamics of quantum systems. To this effect, Shortcuts to Adiabaticity (STA) \cite{Torrontegui2013,Guery2019}, 
stand as a useful toolbox, as they  mimic the result of a slow adiabatic evolution avoiding the drawbacks of long process times, such as decoherence. An open question is to determine their energy cost. 

As devices based on quantum properties, such as quantum computers, engines and refrigerators are being proposed, it 
is  important to understand their 
energy flows and costs.  
A useful model for an engine, exactly solvable but  complex enough to represent friction and heat leaks \cite{Kosloff2017}, considers a quantum harmonic oscillator as the working medium of an Otto engine \cite{Abah2012}. During the cycle, the oscillator undergoes two power strokes between frequencies $\omega_1$ and $\omega_2>\omega_1$, and two thermalizations. Compared to the macroscopic Otto engine, the frequency plays the role of an inverse volume, and the harmonic potential the role of the piston \cite{Kosloff2017}. This quantum engine could be implemented by an ion (primary system) in a Paul trap (control system),  
our model hereafter. Analyzing specific elementary  operations  is important to reach more general conclusions. In particular,  expansions and compressions, apart from being strokes of the Otto engine,  
have been key to develop  STA  in theory \cite{Chen2010a} and experiments \cite{Schaff2010,Schaff2011}.
In a standard analysis the frequency $\omega(t)$ is assumed to be classical, with definite values, whereas the particle is treated quantally. 
This hybrid classical-quantum scenario is justified by the different scales involved and is a general feature, well discussed  in foundational work on quantum mechanics \cite{Bohm1989}, when driving microscopic systems. A related and widespread feature, again a consequence of different scales,  is the negligible effect of the particle on the classical control, whereas the classical control determines the quantum dynamics. As a consequence, we can design useful STA protocols that are independent of the particle state and its dynamics.  

The basic performance criteria of a thermal device are power output and  efficiency. Any thermal device that operates in finite time incurs in rate-dependent losses that diminish efficiency, as opposite to a device operating reversibly with no power output \cite{Andresen1984b}. At the microscopic level, increasing the rate of a given transformation usually increases quantum friction \cite{Kosloff2002,Plastina2014}, i.e., undesired excitations at final time which imply a waste of energy. STA, however, suppress quantum friction  in the power strokes  \cite{Deng2013,DelCampo2014,Abah2018}. It may therefore seem that STA  enhance the power output arbitrarily without affecting efficiency, enabling a  ``perpetual-motion machine of the third kind'' \cite{Salamon2014}. While there is  widespread agreement that some kind of ``cost'' inherent in the STA process precludes these machines, 
many different ``costs'' have been put forward \cite{Chen2010b,Santos2016,Zheng2016,Funo2017,Kosloff2017,Bravetti2017,Abah2018,Calzetta2018}, which are not necessarily in conflict, as long as we 
regard them 
as different aspects of the system energy or its interactions \cite{Guery2019}.

Here, we extend and refine the viewpoint  in \cite{Torrontegui2017, Tobalina2018} and identify the cost with the 
exclusive work for the total system, which is essentially the energy 
consumption to set the driving protocol, i.e. the classical parameters 
of the Hamiltonian for  the primary system. This consumption implies  zero efficiency 
for the quantum Otto engine. 
Several examples dealing with STA demonstrate  that this perspective is crucial to reach sensible conclusions  \cite{Torrontegui2017, Tobalina2018}. Beyond STA see e.g. \cite{Elouard2018, Barra2015,Salamon2014} as examples of the need to account for all the energy flows present in a given process.

%
%
%
%
%
%
%
%
%
%
%
%
%

{\it Definitions of work.} 
We point out two factors  that lead to different definitions for the work done or required by a transformation of a microscopic system. 
The first one corresponds to the definition of work in externally driven systems \cite{Campisi2011}. 
Suppose a simple setting described by the Hamiltonian $H(y,t) = H_0(y)+\lambda(t)y.$ 
with externally driving potential $\lambda(t)y$ along some  coordinate $y$. 
%
The exclusive work definition evaluates only the change of the internal energy defined by the ``unperturbed'' $H_0$. 
It considers this difference as work injected to the unperturbed system by the action of $\lambda(t)$ \cite{Bochkov1977} and corresponds to the standard expression of force times displacement. Instead, an inclusive definition evaluates work as the change of the total Hamiltonian, including the external influence \cite{Horowitz2007}. 


The second factor corresponds to the role of the control.
Gibbs already stated that the force that induces a given transformation on a system is often affected by the configuration of an external body \cite{Gibbs1902}, here the ``control system''. 
The energy needed to manipulate this body should be taken into account when discussing efficiencies, and more so in the context of quantum technologies with  
a macroscopic control system  and a microscopic  ``primary system''. 

Let us combine these factors to propose two useful definitions:  \textit{Total work}  is the exclusive total energy consumed by the 
control plus primary system;  \textit{Microscopic work} is an inclusive definition for the primary system  that disregards the  energy cost to set the control parameters.   The second one is the definition found in most works on quantum thermal machines. It is not invariant under gauge transformations that  shift in time the  zero energy point \cite{Campisi2011}, and thus, according to Cohen-Tannoudji et. al., it is not a physical quantity \cite{Cohen-Tannoudji1991}. The gauge, however, may be fixed  by the experimental setting  to make energy differences physically meaningful \cite{Campisi2011}. Using the proper gauge the microscopic work  may constitute an indicator of the total work if it is proportional to it. 
This proportionality is not guaranteed, 
but it indeed occurs in a specific regime.   
%
%
%
%
%
%
%
%
%
%
%

\begin{figure*}[t]
\includegraphics[width=\textwidth]{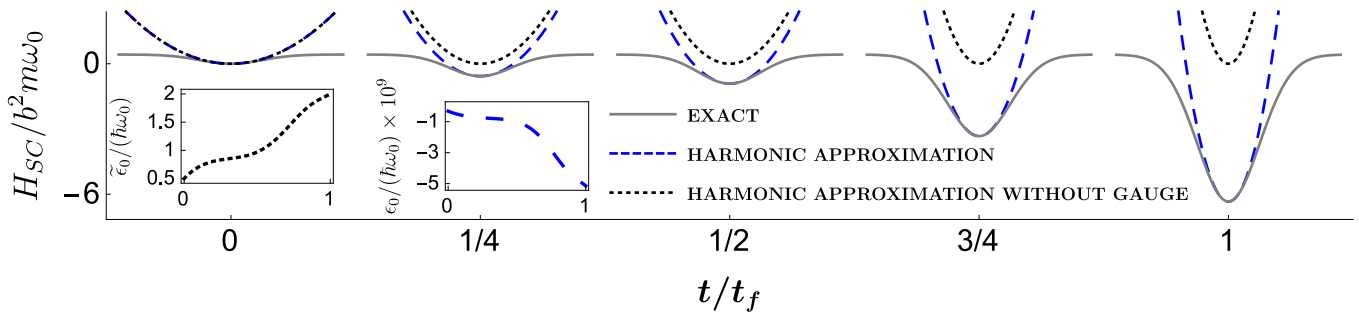}
\vspace*{-.5cm}
\caption{(Color online) Evolution of the interaction  between ion and control for a compression (left to right) or an expansion (right to left).  Exact potential (solid grey line); harmonic approximation in 
Eq. \eqref{intpot} (dashed solid line);  harmonic potential without the gauge term set by the control (dotted black line). The insets depict the time evolution of the  ground energy of the ion in the compression without gauge ($\tilde{\epsilon}_0=\epsilon_0+b^2m\omega^2(t)$) and with gauge term, see Eq. (\ref{stateenergy}). $\omega_0=1.3\times 2\pi$ MHz; $\gamma=1/2$; mass of $^{40}$Ca$^{+}$; $b=0.25\times 10^{-3}$; $t_f=0.2$ $\mu$s. \label{evolutionpotential}}
\end{figure*}

{\it Ion-in-Paul-trap model.} We consider a one dimensional quantum harmonic Otto engine whose working medium is a single ion of mass $m$ and electric charge $Q$. The Hamiltonian for the working medium in the absence of 
the longitudinal harmonic potential reads 
$
H_{S,0}(x)=\frac{p^2}{2m},
$
where $p$ is the momentum of the ion. 
The ion is trapped in a  harmonic potential generated by a segmented linear Paul trap, 
the control system. We model it as a circuit formed by a controllable power source that 
generates  an electromotive force (efm) $\pazocal{E}(t)$ and a low-pass electronic filter formed by a resistor $R$ and a capacitor $C$. Its Hamiltonian is 
\cite{Goldstein2002},  
\beq
H_C(q,t) = q^2 / 2C - \pazocal{E}(t) q,
\eeq
where $q$ is the charge in the capacitor.
The interaction between the control and the ion  is 
$
H_{SC}(x,q,t) = Q \phi(x) q / C,
$
where $x$ is the coordinate  of the ion and $\phi(x)$ is a (dimensionless) electrostatic potential that depends on the geometry of the trap \cite{Furst2014}. Here we use a  Gaussian $\phi(x)=a e^{-x^2/2b^2}$. 
The potential is approximately harmonic near the origin with angular frequency   $\omega(t)$ determined  by   $\partial ^2 H_{SC} / \partial x^2 |_{x=0}= m \omega^2(t)$ and related to $q$ by  
\beq
\label{charge}
 q = - \frac {b^2 m C} { a Q } \omega^2(t).
\eeq
Upon Taylor expansion  around $x=0$,  the interaction  is
\beq
\label{intpot}
H_{SC} = m \omega^2(t)x^2 / 2  - b^2 m \omega^2(t).
\eeq
The last term, usually ignored since it it does not affect the dynamics of the system, fixes the gauge, see Fig. \ref{evolutionpotential}. 

The global Hamiltonian is 
 %
$
H(x,q,t)=\frac{p^2}{2m} + \frac m 2 \omega^2(t)x^2 - b^2 m \omega^2(t) + \frac 1 {2C} q^2 - \pazocal{E}(t) q.
$
%
It governs the evolution of two interacting degrees of freedom, one of them macroscopic and classical, 
the charge in the capacitor, 
and the other microscopic and quantum. 
The overwhelming difference in scale enables  us to make a clear separation and treat the dynamics of the capacitor as effectively independent of the ion dynamics,  whereas the dynamics of the quantum system is governed by 
%
$
H_S=H_{S,0}+H_{SC},
$
%
%
where $\omega(t)$ is treated as an external parameter, whose evolution is designed in what follows using invariant-based inverse engineering.  

This STA technique 
rests on the parameterization of a quadratic invariant 
in terms of a scaling factor $\rho(t)$ that determines  the state width  and satisfies the Ermakov equation \cite{Torrontegui2013,Chen2010a}. 
The evolution of the control parameter is computed from the Ermakov equation as %
$
\omega^2(t)= \omega_0^2 / \rho^4 - \ddot{\rho} / \rho,
$
and $\rho(t)$ is designed to satisfy the boundary conditions $\rho(0)=1$, $\rho(t_f)=\gamma = (\omega_0 / \omega_f)^{1/2}$, $\dot{\rho}(t_b)=0$ and $\ddot{\rho}(t_b)=0$, with $t_b=0,t_f$, so that initial eigenstates of $H_S(0)$ evolve according to 
%
%
\beq
\label{eigenstates}
\psi_n (x,t) = e^{\frac{i m \dot{\rho}}{2 \hbar \rho} x^2} \frac1{\sqrt{\rho}} \phi_n\left(\frac x {\rho} \right),
\eeq
and become at $t_f$ eigenstates of $H_S(t_f)$.
To interpolate  we use
 the ansatz 
$
\rho(t)=\sum_{i=0}^5 \rho_i (t/t_f)^i,
$
with the $\rho_i$ fixed by the boundary conditions. 
We will also use the  notation $\omega_0 \equiv \omega(0)$ and $\omega_f \equiv \omega(t_f)$.
%
%
%
%
%

The expectation value of $H_S$ for the dynamical modes (\ref{eigenstates}) is 
\beq
\label{stateenergy}
\epsilon_n (t)= \frac{(2n + 1) \hbar}{4 \omega_0} \left( \dot{\rho} ^2 + \omega^2(t) \rho^2 + \frac{\omega_0^2}{\rho^2} \right) - b^2 m \omega^2(t),
\eeq
%
see Fig. \ref{evolutionpotential}, and  compare the insets that depict $\widetilde{\epsilon}_0=\epsilon_0+ b^2 m \omega^2(t)$ and  $\epsilon_0$.
The gauge term dominates the evolution of $\epsilon_n(t)$ over the vibrational energy of the ion and, moreover, increases the energy of the ion during the expansion and decreases it during the compression.  

In our inverse engineering protocol, once the desired $\omega(t)$ has been set, the dynamics of the capacitor charge $q$  is found from  Eq. \eqref{charge}. Using the modified Hamilton equation $\dot p_q = - \frac{\partial H_C}{\partial q} -  \frac{\partial \pazocal F}{\partial \dot q} $, which accounts for friction through  Rayleigh's dissipation function \cite{Goldstein2002} $\pazocal F = R \dot q ^2 /2$, we get  the electromotive force  to carry out the designed process,
\beq
\label{electromotiveforce}
\pazocal{E} (t)= \frac q C + R \dot q.
\eeq
%
%
%
%
%
%
%
%
%
%
%

%
{\it Work and power.} Work is commonly computed as  $W = \int_0^t \pazocal P dt'$, where $\pazocal P$ represents instantaneous power. Total work considers the evolution of the exclusive instantaneous power of the composite unperturbed system (defined by $H_0=H+ \pazocal{E}(t) q$).
To estimate its effect we consider 
a classical approximation of the microscopic system.  Later we shall substitute  
the variables dealing with the quadratic microscopic system by quantum expectation values.  
For systems described by a Rayleigh dissipation function, the modified 
Hamilton equations imply that  the (exclusive) total power contributes to change the 
energy of the unperturbed system and to overcome friction, see Suppl. Material. In our model,
\beq
\label{powergeneral}
\pazocal P = \pazocal{E}(t) \dot{q}= dH_0 / dt + R \dot q ^2. 
\eeq
%
We can separate it  as $\pazocal P = \pazocal P_C + \pazocal P_S$, the power needed to generate the dynamics on the Paul trap (without the ion) and the power  to overcome the backaction by the ion, 
\beqa
\label {pc}
\pazocal P_C &=& \frac{\partial H_{0,C}}{\partial t} + R \dot q ^2 = (R \dot q + q/C ) \dot q,
\\
\label{ps}
\pazocal P_S &=& \frac{\partial H_S}{\partial t}  =  \left(1- \frac {x^2} {2 b^2}\right)\frac{aQ}C \dot q, 
\eeqa
where $H_{0,C}=q^2 / 2C$. 
In $\pazocal P_S$ the first term is due to the gauge term. 
As $q>>Q$ we expect  $|\pazocal P_S| \ll |\pazocal P_C|$, and  $\pazocal P \approx \pazocal P_C = \pazocal E(t) \dot q$, 
see Eq. (\ref{electromotiveforce}), which holds in all calculations. 
The dominance of $\pazocal P_C$ is needed to set  state-independent  shortcut protocols. 

%
The possibility to ``regenerate'' (store and reuse) the energy that flows out of the system during negative-power time segments  is accounted for by a factor $-1\le \mu \le1$  \cite{Torrontegui2017} multiplying the negative power  in the total integrated work,   
\beq
\label{workmu}
W_T = \int_0^t \pazocal P_{C_+} dt' + \mu \int_0^t \pazocal P_{C_-} dt'.
\eeq
Here  $\pazocal P _{C_{\pm}} = \Theta(\pm \pazocal P_C) \pazocal P_C$, and $\Theta$ is the Heaviside function. In the Paul trap both signs require consumption, i.e.,  $\mu = -1$.  

The backaction term is in fact the inclusive microscopic work. The quantum version of Eq. (\ref{ps}) takes the same form with expectation values,  $H_S\to\langle H_S\rangle$, $x^2\to\langle x^2\rangle$.  
Defining work at the quantum level is not straightforward  \cite {Talkner2007} and, furthermore, the relation between the system energy and inclusive work holds only if the gauge is appropriately fixed according to the experiment \cite{Vilar2008,Campisi2011}. We evaluate the microscopic work for the shortcut process as the difference between final and initial energies determined by $H_S$.   
The contribution from each mode is 
%
$
\langle W_m \rangle = \sum_n p_n^0 \int_0^{t_f} \langle \pazocal P_S \rangle_n dt,
$
%
%
where 
$
\langle \pazocal P_S \rangle_n = d \epsilon_n/dt=(2n+1)\hbar \rho^2\omega\dot\omega/(2\omega_0) - 2 b^2 m \omega \dot \omega.
$
%
Assuming an initial thermal distribution with  $p_n^0 = e^{-\beta \epsilon_n(0)}/Z$,  inverse temperature $\beta$, and $Z=\sum_n e^{-\beta \epsilon_n(0)}$, we get 
%
%
%
%
\beq
\label{work1}
\langle W_m \rangle = \frac{\hbar}2 (\omega_f - \omega_0) \coth \left( \frac{\beta \hbar \omega_0} 2 \right) - b^2 m (\omega_f ^2- \omega_0 ^2).
\eeq
%
Importantly, for any realistic $\beta$, $\langle W_m \rangle$ is negative when $\omega_0 > \omega_f$ and positive when $\omega_0 < \omega_f$. Owing to the gauge term, the microscopic work in each power stroke behaves  opposite of what it 
is commonly expected when ignoring the physical gauge, see again the inset of Fig. \ref{evolutionpotential}.


%
%



%
%
%
%
%
%
%
%
%
%
%
%
%
%

\begin{figure}
\begin{center}
\includegraphics[width=1.04\linewidth]{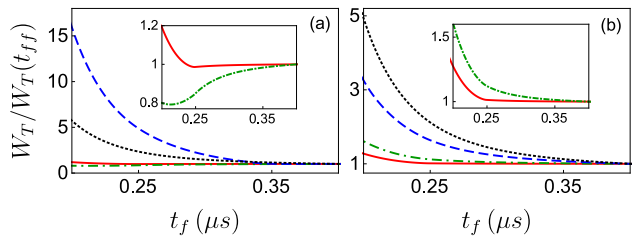}
\caption{(Color online) Normalized total work for  a) an expansion or b) a compression  between $\omega_1= 2\pi$ MHz and $\omega_2 = 2 \times 2\pi$ MHz versus final times. $W_T$ is computed with Eq. (\ref{workmu}). 
For all curves ${\pazocal P}_C$ is positive for expansions and negative for 
compressions. The normalization constant is the work for  a process time $t_{ff} = 0.4$ $\mu$s. Different curves correspond to:   $R=3$ $\Omega$, $C = 1$ nF (red solid line);  $R=3$ $\Omega$,  $C = 10$ nF (green dash-dotted line);  $R=300$ $\Omega$, $C = 1$ nF (blue dashed line); $R=300$ $\Omega$, $C = 10$ nF (black dotted line). The insets zoom in the first two lines. \label{WT_vs_tf}}
\end{center}
\end{figure}

{\it Cost of STA.} The total work in Eq. \eqref{workmu} comes from the two terms of $\pazocal P_C$ in Eq. \eqref{pc}, an Ohmic dissipation  $R\dot{q}^2$ and the change in the potential energy of the capacitor. 
Using the time constant of the  circuit ($RC$) and  $q/\dot{q}=\omega/ (2 \dot \omega)$ we identify different regimes, dominated  by the dissipative term or by the capacitor term.
When $\omega/ (2 \dot \omega) \ll RC$, $\pazocal P_C$ is dominated by dissipation in the resistor, and, 
compare Eqs. (\ref{pc}) and (\ref{ps}) and note the dominance of the gauge (first term) in Eq. (\ref{ps}),    
%
$
\pazocal P_C / \dot q \approx  RC \, \pazocal P_S / (a Q).
$
%
Alternatively, when $\omega/ (2 \dot \omega) \gg RC$,  the instantaneous microscopic power  per unit charge is proportional to the control power input per unit charge, 
%
$
\pazocal P_C / q \approx  \pazocal P_S  / (a Q).
$
%
Figure \ref{WT_vs_tf} shows the total work consumed by the power strokes in  different regimes, for final times that yield a monotonic  $\omega(t)$. For an expansion stroke dominated by the capacitor, the consumption decreases as we reduce $t_f$  up to $t_f\approx 0.21$ $\mu$s. 
For faster protocols, $\omega/ (2 \dot \omega) \gg RC$ becomes unattainable and  total work is dominated by dissipation. We check numerically that for both power strokes, as $t_f\rightarrow0$, the total work scales as $1/t_f^{\,\,\,5}$. This scaling agrees with 
\cite{Torrontegui2017,Tobalina2018} and contrasts with Landauer's estimate of the energy dissipation as being proportional to the ``velocity of the process'' when studying the cost of computation \cite{Landauer1987}. 

%
%
%
%
%
%
%
%
%
%
{\it Performance of the engine.} The 
microscopic power output of the engine is calculated along a cycle time $\tau$, adding compression and expansion 
terms,
%
$
P = {(\langle W_m^{\rm comp} \rangle - \langle W_m^{\rm exp} \rangle)}/ \tau.
$
%
STA protocols increase this quantity by reducing $\tau$ and keeping the work output of adiabatic processes.  

Typically, accelerating the thermodynamic cycle increases dissipation, diminishing work output and/or rising the energy required to perform the cycle, reducing efficiency. 
For a typical engine the input energy is the heat absorbed from the hot bath, $\langle {\pazocal{Q}} \rangle$. Since the shortcut does not affect the heat absorbed, it may seem that the efficiency  is not  affected.
STA engines, however, are not typical engines, as the expansion of the piston (in our case the trap expansion) is externally driven rather than being a consequence of pushing by  a hot working medium. The compression is similarly externally driven
with a cost. Thus, the energy used to generate the external driving in Eq. \eqref{workmu} constitutes an extra energy demand.
The efficiency becomes 
\beq
\label{efficiency}
\eta = \frac{\langle W_m^{\rm comp} \rangle - \langle W_m^{\rm exp} \rangle}{\langle {\pazocal{Q}} \rangle + W_T^{\rm exp} + W_T^{\rm comp}},
\eeq
which is  negligible, $\eta \approx 0$, due to the scale difference between the working medium and the control.
{Notice that to calculate the engine performance criteria we use the microscopic work. 
Along the cycle the gauge contributions to the microscopic work in each power stroke exactly cancel each other, making engine power and efficiency truly physical quantities whether we use or do not use  the ``physical gauge'' provided by the experiment. This effect is similar to what happens with the Jarzynski inequality \cite{Jarzynski1997}, also given in terms of inclusive work.}

The idea of finding the optimal time path for the motion  of the piston has been 
present in the field of Finite-Time Thermodynamics for long \cite{Mozurkewich1981}. However, already Andresen et. al. pointed out that such trajectory may be optimal for some performance criterion but detrimental for others \cite{Andresen1984b}, as we see here in a clear and extreme way.

{\it Conclusions.} 
STA are externally driven processes. Even if they evade quantum friction in the microscopic system, they need resources to drive the macroscopic control, more so for shorter process times, due to  
dissipation in the macroscopic control. 

Once the cost of the shortcut is taken into account, the efficiency of an Otto-cycle engine for an ion in a Paul trap goes to zero. 
This challenges scalability, in other words, the idea that
a device that combines many STA quantum engines may outperform a classical engine. 
A quantum and macroscopic primary system is a route to explore \cite{Guery2019}, 
but it could make STA state dependent.   

One way to decrease the total consumption is ``regeneration'', 
i. e., achieving a $\mu \approx 1$ regime with smart designs,  
but even if some energy is regenerated, the fast nature of STA will imply dissipation in the control. 
A 
cornerstone of finite-time thermodynamics is that processes without dissipation do not occur in nature if performed in finite time.     

There are quantum engines that replicate the behavior of typical engines, the so called \textit{autonomous engines}. In contrast with STA engines, the power strokes of an autonomous engine are not externally driven, so there is no need of power input to control them. An experimental implementation was presented in \cite{Rossnagel2016}. 
In principle STA are not applicable in autonomous engines.
There exists however, the possibility to accelerate the other two strokes of the cycle, the thermalization processes, in order to achieve shorter times and increase power output 
\cite{Martinez2016,Dann2018,Villazon2019},
but they may involve  extra energy consumption that should be studied. 

\acknowledgments{This work was supported by the Basque Country Government (Grant No. IT986-16)
and PGC2018-101355-B-I00 (MCIU/AEI/FEDER,UE).}

\bibliography{Bibliography}{}
\bibliographystyle{sofia} 
\appendix
\section{Supplementary material: Exclusive power by external force} \label{proof}
Here we prove that the expression in Eq. \eqref{powergeneral} can be generalized to any system whose dissipation is proportional to the velocity squared, and thus describable by a Hamiltonian complemented by a Rayleigh dissipation function $\pazocal F = \frac 1 2 \gamma \dot q^2$. We do it for one dimension, but it can be extended to higher dimensions. We start by separating the total Hamiltonian into the unperturbed system and the external, time-dependent force term, $H(q,p_q,t)= H_0(q,p_q) - F(t) q$. The rate of change of the unperturbed Hamiltonian then reads 
\beq
\frac{dH_0}{dt} = \frac{dH}{dt} + \dot F q + F \dot q.
\eeq
Rewriting the time derivative of the total Hamiltonian,
\beq
\frac{dH}{dt} = \frac{\partial H}{\partial t} + \frac{\partial H}{\partial q} \dot q + \frac{\partial H}{\partial p} \dot p,
\eeq
and then using modified Hamilton equations, $\dot q = \partial H / \partial p_q$ and $\dot p_q = - \partial H / \partial q - \partial \pazocal F / \partial \dot q$, we get 
\beq
\frac{dH_0}{dt} = \frac{\partial H}{\partial t} - \dot q \frac{\partial \pazocal F }{ \partial \dot q}  + \dot F q + F \dot q.
\eeq 
Notice that the only explicitly time dependent element of the Hamiltonian is the external force, and thus $\partial H / \partial t = - \dot F q $. Finally, performing the derivative of the Rayleigh function and reordering the terms, we get
\beq
\frac{dH_0}{dt} + \gamma \dot q ^2 = F \dot q.  
\eeq

\end{document}